\documentclass{article}
\usepackage{spconf,amsmath,graphicx,hyperref}
\usepackage{amsfonts}
\usepackage{amsmath}
\usepackage{xcolor}

\usepackage{caption}
\title{A Novel Monte Carlo Gradient Method Based on Meta-learning for Effective Step-size Selection in Active Noise Control}
\name{
Luyuan Li{\rm\textsuperscript{1}}, Jisheng Bai{\rm\textsuperscript{2}}, Xiruo Su{\rm\textsuperscript{3}}, Xiaoyi Shen{\rm\textsuperscript{4}}, Dongyuan Shi{\rm\textsuperscript{1}}, Woon-seng Gan{\rm\textsuperscript{5}}
}

\address{
	\textsuperscript{1} Center of Intelligent Acoustics and Immersive Communications\\
    Northwestern Polytechnical University, Xi'an, China\\
    \textsuperscript{2} School of Communications and Information Engineering, Xi'an University of\\Posts and Telecommunications, Xi'an, China\\
    \textsuperscript{3}Zhejiang Provincial Key Laboratory for Network Multimedia Technology,\\ Zhejiang University, Hangzhou, China\\
    \textsuperscript{4}State Key Laboratory of Acoustics and Marine Information, Institute of Acoustics, \\Chinese Academy of Sciences, Beijing, China\\
    \textsuperscript{5}School of Electrical and Electronic Engineering, Nanyang Technological University, Singapore\\
}

\begin{document}
%\ninept
%
\maketitle
\begin{abstract}
% Active noise control (ANC) is an effective method for noise suppression, and the filtered-reference least mean square (FxLMS) algorithm is a widely used technique in ANC systems due to its computational efficiency and stable performance. However, its convergence speed and noise reduction effects are highly dependent on the step size parameter. Common step-size algorithms, such as normalized and variable step sizes, demand additional computational resources and suffer from limited adaptability across varying environments. To address this issue, a novel Monte Carlo gradient meta-learning (MCGM) approach is proposed to determine an appropriate step size, with a forgetting factor incorporated to mitigate the effect of initial zeros. Compared with other algorithms, the proposed method does not introduce any additional computational burden to the FxLMS operation. Numerical simulations with real-world paths and noise further confirm its effectiveness and robustness.

Active noise control (ANC) is an effective approach to noise suppression, and the filtered-reference least mean square (FxLMS) algorithm is a widely adopted method in ANC systems, owing to its computational efficiency and stable performance. However, its convergence speed and noise reduction performance are highly dependent on the step size parameter. Common step-size algorithms—such as normalized and variable step-size variants—require additional computational resources and exhibit limited adaptability under varying environmental conditions. To address this challenge, a novel Monte Carlo gradient meta-learning (MCGM) approach is proposed herein to determine an appropriate step size, into which a forgetting factor is incorporated to mitigate the impact of initial zero effect. Compared to other algorithms, the proposed method imposes no additional computational burden on FxLMS operations. Numerical simulations involving real-world acoustic paths and noise signals further confirm its effectiveness and robustness.
\end{abstract}
\begin{keywords}
Adaptive noise control (ANC), meta-learning, step size, fast convergence.
\end{keywords}
\section{Introduction}
\label{sec:intro}

Acoustic noise problems are gaining growing attention with the increased number of industrial equipment in daily life~\cite{uds_ANC}. Active noise control (ANC) is a technology designed to address this issue by generating a destructive anti-noise wave that reduces unwanted sound at a specific location~\cite{elliot1994active, kuo2002active}. Unlike passive solutions, such as noise barriers, ANC is particularly effective in mitigating low-frequency noise, making it ideal for applications in headphones, vehicles, windows, ventilation systems, etc.~\cite{lei2015active,shen2022adaptive}.

The filtered-reference least mean square (FxLMS) algorithm plays an important role in modern ANC systems due to its computational efficiency and reliable steady-state noise attenuation capability~\cite{ardekani2010theoretical,gaur2016review}. However, its convergence speed is constrained by the step size. A larger step size accelerates convergence but introduces the risk of divergence, whereas a smaller step size offers greater stability at the cost of slower convergence~\cite{miyake2023head, shi2020feedforward}. To address this, variable step-size approaches have been proposed to automatically adjust the step size, accelerating convergence while maintaining stability. Among these algorithms, the step-size parameter is commonly formulated as a function of the error signal~\cite{huang2012variable}, the reference signal~\cite{jiang2022laboratory}, or both~\cite{yin2023adaptive, liu2025robust}. Nevertheless, the step-size functions in these algorithms are typically empirical and remain without comprehensive theoretical underpinning, which limits their generalization and effectiveness in diverse noise environments. Additionally, the combined step-size method~\cite{akhtar2023developing, zhou2024combined} utilizes two distinct step-size parameters with dynamically adjusted weighting coefficients to balance convergence speed and steady-state misalignment. A golden-section search~\cite{shi2023computation} has been employed to preselect a fast step size before noise cancellation. However, such strategies are generally tailored to specific noise types and tend to degrade performance when applied to different noise environments.

Recently, some studies~\cite{shi2022selective,luo2022hybrid} have introduced machine learning into the ANC system to improve convergence speed. Among these strategies, meta-learning~\cite{finn2017model,casebeer2023meta} is a highly effective algorithm for determining optimal parameters, and some scholars have already applied it to ANC system~\cite{ shi2024behind, shen2025data}. Based on meta-learning theory, this paper proposes a novel Monte Carlo gradient method to train and obtain an optimal step size, while introduces a forgetting factor to mitigate the initial zero effect during filter initialization. The detailed theory and numerical experiments in the paper demonstrate that the step size selected by this method will effectively accelerate the convergence of FxLMS and achieve superior noise reduction performance.

\section{Proposed method}
\label{sec:method}

\subsection{Filtered Reference Least Mean Square Algorithm}

\begin{figure}[t]
    \centering
    \includegraphics[width=0.9\linewidth]{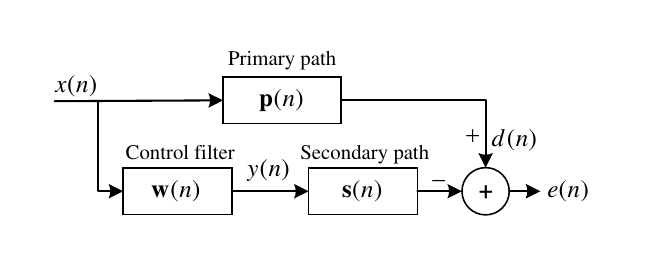}
    \caption{Block diagram of a feedforward active noise control system}
    \label{fig:ff_ANC}
\end{figure}

This paper focuses on a feedforward ANC system, as presented in Fig. \ref{fig:ff_ANC}.  The error signal $e(n)$ can be expressed as the difference between the disturbance signal $d(n)$ and the anti-noise $y(n)*s(n)$:
%---------------------------------------------------------------------------------------
\begin{equation}
    e(n)=d(n)-y(n)*s(n) \label{e(n)},
\end{equation}
%---------------------------------------------------------------------------------------
where $*$ indicates the convolution operation, $s(n)$ represents the secondary path, and  $y(n)$ denotes the control signal, which is generated by the control filter as follows
%---------------------------------------------------------------------------------------
\begin{equation}
    y(n)=\mathbf{x}^{\mathrm{T}}(n) \mathbf{w}(n) \label{y(n)}.
\end{equation}
%---------------------------------------------------------------------------------------
Here, $\mathbf{x}(n)=[x(n), x(n-1), \cdots, x(n-N+1)]^{\mathrm{T}}$ and $\mathbf{w}(n)=[w_1(n),w_2(n),\cdots,w_{N}(n)]^{\mathrm{T}}$ denotes the reference signal vector and the coefficient vector of control filter, both of length $N$. $\mathrm{T}$ indicates the transpose operation.

%--------------------------------------------------------------------------------------
%--------------------------------------------------------------------------------------

To minimize the mean square error defined in \eqref{e(n)}, the filtered-x least mean square (FxLMS) algorithm is employed to iteratively update the control filter as
%---------------------------------------------------------------------------------------
\begin{equation}\label{FxLMS_UPATE}
\mathbf{w}(n+1)=\mathbf{w}(n)+\mu e(n)\mathbf{x}^{\prime}(n),
\end{equation}
%---------------------------------------------------------------------------------------
where $\mathbf{x}^{\prime}(n)=\left[x^{\prime}(n),x^{\prime}(n-1),\cdots x^{\prime}(n-N+1)\right]^{\mathrm{T}}  \label{X'(n)}$ denotes the filtered reference signal vector, which is given by 
%---------------------------------------------------------------------------------------
\begin{equation}\label{X'(n)}
\mathbf{x}^{\prime}(n)=\mathbf{x}(n)*s(n). 
\end{equation}
%---------------------------------------------------------------------------------------

In \eqref{FxLMS_UPATE}, $\mu$ denotes the step size, which governs the convergence behavior of the FxLMS algorithm. An improperly chosen step size may cause instability or significantly slow convergence. In conventional operations, a theoretical value can be obtained from 
%---------------------------------------------------------------------------------------
\begin{equation}
    \mu_{\text{c}}=\frac{1}{P_{x}L}, \label{mu_c}
\end{equation}
%---------------------------------------------------------------------------------------
where $P_{x}$ and $L$ denote the power of the filtered reference signal and the delay of the secondary path estimate, respectively. However, the step size obtained through \eqref{mu_c} still requires empirical tuning to enhance the noise reduction performance.

\subsection{Monte-Carlo Gradient Meta-learning algorithm}

This paper introduces a Monte Carlo gradient meta-learning (MCGM) algorithm to determine an appropriate step size for the FxLMS algorithm, enabling faster convergence without relying on trial-and-error. 

Since the ANC noise distribution approximates a uniform mixture model (UMM)~\cite{ shi2024behind}, Monte Carlo sampling is adopted to generate the meta-learning dataset. Specifically, we begin with randomly sampling $K$ sets of noise data from the noise environment, including both reference signal vectors and disturbance signal vectors, denoted as $\mathbf{x}^{(k)} = [x^{(k)}(N-1), \cdots, x^{(k)}(1), x^{(k)}(0)]^{\mathrm{T}}$ and $\mathbf{d}^{(k)} = [d^{(k)}(N-1), \cdots, d^{(k)}(1), d^{(k)}(0)]^{\mathrm{T}}$, respectively. And the corresponding filtered reference signal, $\mathbf{x}^{\prime(k)} = [x^{\prime(k)}(N-1), \cdots, x^{\prime(k)}(1), x^{\prime(k)}(0)]^{\mathrm{T}}$, is obtained from \eqref{X'(n)}. Here, $k=0,1,\cdots,K-1$.

%---------------------------------------------------------------------------------------
\begin{figure}[t]
    \centering
    \includegraphics[width=0.96\linewidth]{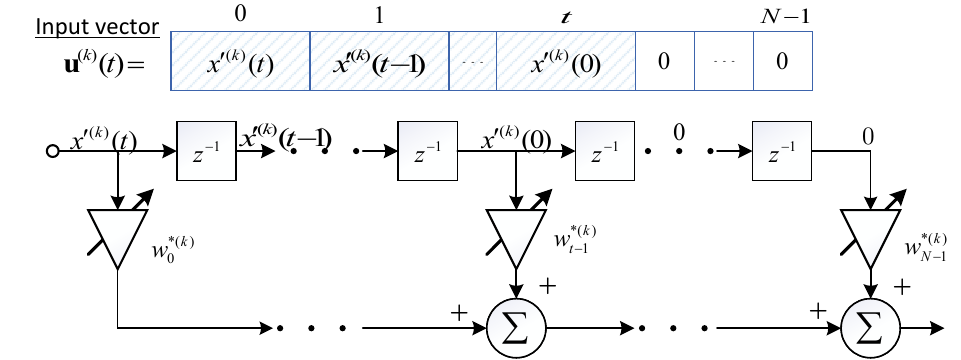}
    \caption{Equivalent block diagram of FxLMS at the initial $t$th iteration}
    \label{fig:N-t diagram}
\end{figure}
%---------------------------------------------------------------------------------------

 To simulate the initial state of the input delay line, we construct $N$ input vectors for each task, as shown in Fig. \ref{fig:N-t diagram}. The $t$th ($t=0,1, \cdots,N-1$) input vector to the control filter during $k$th task and is defined as
%---------------------------------------------------------------------------------------
\begin{equation} 
        \mathbf{u}^{(k)}(t)=[x^{\prime(k)}(t),\cdots,x^{\prime(k)}(0), \mathbf{0}_{1\times {(N-1-t)}}]^\mathrm{T}. \label{u(t)}
\end{equation}
%---------------------------------------------------------------------------------------
Then, the error signal $e^{(k)}(t)$ can be expressed as
%---------------------------------------------------------------------------------------
\begin{equation}   
        e^{(k)}(t)=d^{(k)}(t)-[\mathbf{u}^{(k)}(t)]^\mathrm{T}\mathbf{w}^{(k)}(t)  \label{e(t)}. 
\end{equation}
%---------------------------------------------------------------------------------------
and the iterative formula for control filter is derived from \eqref{FxLMS_UPATE}, \eqref{u(t)}, and \eqref{e(t)} as
%---------------------------------------------------------------------------------------
\begin{equation}
 \mathbf{w}^{(k)}(t+1)= \mathbf{w}^{(k)}(t)+\mu^{(k)}e^{(k)}(t)\mathbf{u}^{(k)}(t).  \label{w(t)}
\end{equation}
%---------------------------------------------------------------------------------------

Following this, we can obtain the corresponding cost function from
%---------------------------------------------------------------------------------------
\begin{equation} 
    \mathbb{J}^{(k)}(t)=\left[e^{(k)}(t)\right]^2 \label{J(n)}.
\end{equation}
%---------------------------------------------------------------------------------------

%---------------------------------------------------------------------------------------
\begin{table}[t!] 
    \centering
    \caption{Pseudo-code of MCGM-based step-size method}\label{Pseudocode}
    \begin{tabular}{l}
    \hline
    \textbf{Algorithm:} MCGM-based step-size method\\
    \hline
    \textbf{Initializing} the epoch number $K$, the forgetting factor $\lambda$,\\ and the learning rate $\alpha$; Setting iteration index $k$ to $0$.\\ 
    1:~\textbf{while} $k \neq K$  \textbf{do}\\
    2:~~~~\textbf{Randomly sampling:} the disturbance signal $\mathbf{d}$, the\\
    ~~~~~~~reference signal $\mathbf{x}$ and its corresponding filtered ref-\\
    ~~~~~~~erence signal $\mathbf{x}^{\prime}$.\\
    3:~~~~\textbf{Constructing} input vectors: $[\mathbf{u}(0)$, $\cdots$, $\mathbf{u}(N-1)]$.\\
    4:~~~~\textbf{Initializing} the control filter:  $\mathbf{w}(0)=[0\cdots0]$.\\ 
    5:~~~~\textbf{for} $t$ \textbf{in} $[0,1,\cdots,N-1]$ \textbf{do}\\
    6:~~~~~~~~$e(t) \leftarrow d^{}(t)-[\mathbf{u}(t)]^\mathrm{T}\mathbf{w}(t)$,\\
    7:~~~~~~~~$\mathbf{w}(t+1)=\mathbf{w}(t)+\mu e(t)\mathbf{u}(t)$.\\
    8:~~~~\textbf{end~for}\\
    9:~~~~\textbf{Updating:} $\mu \leftarrow \mu+\alpha\sum_{t=1}^{N-1}\lambda^{(N-1-t)} e(t) 
      [\mathbf{u}(t)]^\mathrm{T}$\\
    ~~~~~~~~~~~~~~~~~~~~~~~~~~~~~~~~~~$\times \left[ \sum_{i=1}^t e(i-1)\mathbf{u}(i-1) \right]$.\\
    10:~~~$k \gets k+1$.\\
    11:~\textbf{end~while}\\
    \hline 
    \end{tabular}
\end{table}

%---------------------------------------------------------------------------------------

The loss function during the $k$th task is defined as a weighted sum of the cost function over $N$ iterations
%---------------------------------------------------------------------------------------
\begin{equation}
\mathbb{L}^{(k)}(\mu)=\sum_{t=0}^{N-1}\lambda^{(N-1-t)}\mathbb{J}^{(\mathrm{k})}(t) \label{L(phi)}.
\end{equation}
%---------------------------------------------------------------------------------------
where $\lambda$ ($0 < \lambda < 1$) is the forgetting factor, which is introduced to mitigate the impact of initial zeros.

Since \eqref{L(phi)} is a convex function with respect to $\mu$, we can apply the gradient descent algorithm to reach the global minimum. Thus, the recursive update formula for the step size is given by
%---------------------------------------------------------------------------------------
\begin{equation}\label{miu_update}
    \mu^{(k+1)}=\mu^{(k)}-\frac{\alpha}{2}\frac{\partial \mathbb{L}^{(k)}(\mu)}{\partial\mu},
\end{equation}
%---------------------------------------------------------------------------------------
where $\alpha$ ($0<\alpha<1$) stands for the learning rate.

The gradient of $\mathbb{L}^{(k)}(\mu)$ in terms of $\mu$ is deduced from \eqref{e(t)}, \eqref{J(n)}, and \eqref{L(phi)} as
%---------------------------------------------------------------------------------------
\begin{equation}
    \begin{aligned}
    &\frac{\partial \mathbb{L}^{(k)}(\mu)}{\partial\mu}=2\sum_{t=0}^{N-1}\lambda^{(N-1-t)}\mathrm{e}^{(k)}(t)\frac{\partial\mathrm{e}^{(k)}(t)}{\partial\mu} \\
     &=-2\sum_{t=0}^{N-1}\lambda^{(N-1-t)}\mathrm{e}^{(k)}(t)[\mathbf{u}^{(k)}(t)]^\mathrm{T}\frac{\partial\mathbf{w}^{(k)}(t)}{\partial\mu}.    \label{L(k)pd}
    \end{aligned}
\end{equation}
%---------------------------------------------------------------------------------------
According to \eqref{w(t)}, the partial derivative of $\mathbf{w}^{(k)}(t)$ with respect to $\mu$ is given by
%---------------------------------------------------------------------------------------
\begin{equation}
\begin{aligned}
\frac{\partial\mathbf{w}^{(k)}(t)}{\partial\mu}=
\begin{cases}
0, & t=0 \\
\sum_{i=1}^te^{(k)}(i-1)\mathbf{u}^{(k)}(i-1), & t>0  \label{w*pd}
\end{cases}
\end{aligned}
\end{equation}
%---------------------------------------------------------------------------------------

%---------------------------------------------------------------------------------------
\begin{figure}[t!]
    \centering
    \includegraphics[width=0.93\linewidth]{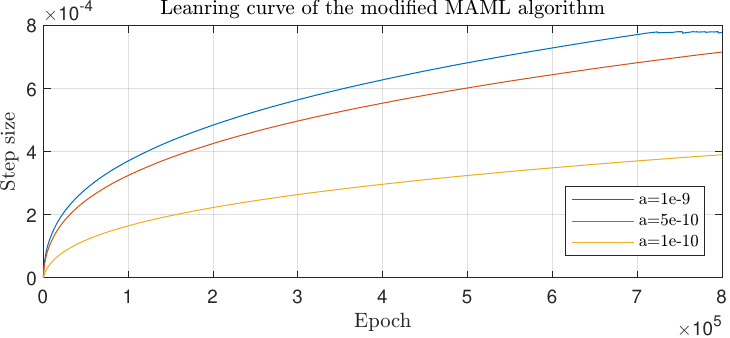}
    \caption{Learned step size over time with different $\alpha$ values.}
    \label{fig:learncurve}
\end{figure}
%---------------------------------------------------------------------------------------

Hence, substituting \eqref{w*pd} into \eqref{L(k)pd} yields
%---------------------------------------------------------------------------------------
\begin{equation}\label{loss_gradient}
\begin{split}
\frac{\partial \mathbb{L}^{(k)}(\mu)}{\partial\mu} 
= & -2\sum_{t=1}^{N-1}\lambda^{(N-1-t)} e^{(k)}(t) 
      [\mathbf{u}^{(k)}(t)]^\mathrm{T} \\
& \times \left[ \sum_{i=1}^t e^{(k)}(i-1)\mathbf{u}^{(k)}(i-1) \right].
\end{split}
\end{equation}
%---------------------------------------------------------------------------------------
Finally, substitute \eqref{loss_gradient} into \eqref{miu_update} as
%---------------------------------------------------------------------------------------
% \begin{equation}\label{gradient_descent}
%     \begin{aligned}
%     &\mu^{(k+1)}=\mu^{(k)}+\\
%     &\alpha\sum_{t=0}^{N-1}\lambda^{(N-1-t)}\mathrm{e}^{{(k)}}(t)e^{(k)}(t-1)[\mathbf{u}^{(k)}(t)]^\mathrm{T}\mathbf{u}^{(k)}(t-1).
%     \end{aligned}
% \end{equation}
\begin{equation}\label{gradient_descent}
    \begin{aligned}
    \mu^{(k+1)}=&\mu^{(k)}+\alpha\sum_{t=1}^{N-1}\lambda^{(N-1-t)} e^{(k)}(t) 
      [\mathbf{u}^{(k)}(t)]^\mathrm{T}\\
    &\times \left[ \sum_{i=1}^t e^{(k)}(i-1)\mathbf{u}^{(k)}(i-1) \right].
    \end{aligned}
\end{equation}
%---------------------------------------------------------------------------------------

By employing \eqref{gradient_descent}, the iteration for $K$ tasks is completed, yielding the optimal step-size for the dataset. It is noteworthy that these training procedures are completed prior to the execution of the FxLMS algorithm, thus introducing no additional computational complexity during its operation. The detailed pseudocode is presented in Table ~\ref{Pseudocode}.

\section{Numeral Simulations}\label{sec:Experiments}

This section presents a series of simulation experiments to assess the performance of the proposed algorithm. The primary and secondary paths were measured from an actual noise duct, with impulse response lengths of 512 and 256, respectively, while the control filter order was set to 512. The noise data, including real-world recorded noise and synthetic noise, were partitioned into $70\%$ for training and $30\%$ for testing. All experiments were conducted within the FxLMS algorithm, where the proposed MCGM step-size method was compared with other strategies, including the constant theoretical step size (from \eqref{mu_c}), normalized step size~\cite{uds_ANC}, variable step size~\cite{yin2023adaptive}, and combined step size~\cite{akhtar2023developing}.

\subsection{Broadband noise cancellation}
\label{sec:Broadband}
In this simulation, the noise dataset comprises four broadband noises, whose frequency bands span from $0.6-1.8$ kHz, $1.5-4.0$ kHz, $3.5-5.0$ kHz, and $4.4-6.0$ kHz,  and system sampling rate is set to $16$ kHz. Fig.~\ref{fig:learncurve} illustrates the learning curves of the step size under different learning rates $\alpha$, with the forgetting factor fixed at 0.5. It can be observed that the magnitude of $\alpha$ significantly influences the convergence speed. Consequently, we set $\alpha = 1 \times 10^{-9}$ for the experiments in this study.

\begin{figure}[t]
    \centering
    \includegraphics[width=0.93\linewidth]{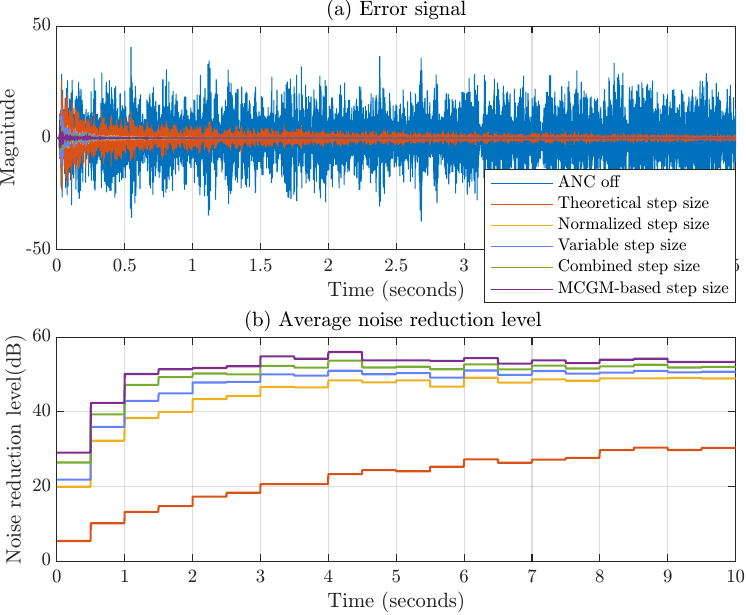}
    \caption{Broadband noise reduction performance of FxLMS with different step-size strategies: (a) the error signal and (b) the average noise reduction level in every 0.5 s.}
    \label{fig:broadband}
\end{figure}

The broadband noise reduction performance is shown in Fig.~\ref{fig:broadband}, including the error signal and the average noise reduction level. The results indicate that incorporating the proposed MCGM-based step size into the FxLMS algorithm leads to faster convergence and achieves superior noise reduction. By comparison, the theoretical step size yields the poorest performance, whereas the normalized, variable, and combined step-size methods achieve relatively better results.

%---------------------------------------------------------------------------------------
%---------------------------------------------------------------------------------------

\subsection{Real-world noise cancellation}
\begin{figure}[t]
    \centering
    \includegraphics[width=0.91\linewidth]{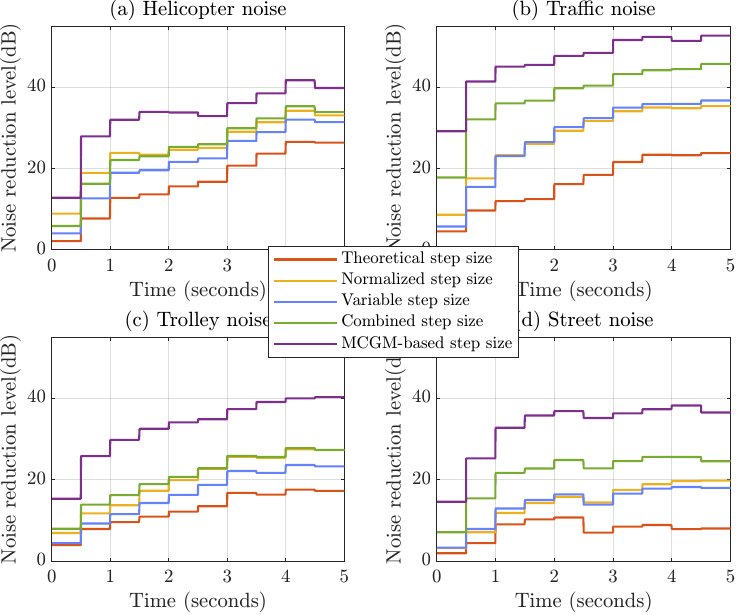}
    \caption{Average noise reduction level in every 0.5 s for various real-world noise, using FxLMS with different step-size strategies.}
    \label{fig:realnoise}
\end{figure}

This simulation investigates the effectiveness of the proposed method under various real-word noises, including helicopter noise, traffic noise, trolley noise, and street noise. And the training set is composed of four noise types in equal proportions.
Fig.~\ref{fig:realnoise} illustrates the average noise reduction level in every 0.5 s for various real-world noises. It can be observed that the MCGM-based step size yields superior noise suppression results for all four noise conditions. However, some methods demonstrate effective noise reduction performance only under specific noise types. 
%---------------------------------------------------------------------------------------
%---------------------------------------------------------------------------------------

\subsection{Dealing with the varying secondary path}
\begin{figure}[t]
    \centering
    \includegraphics[width=0.86\linewidth]{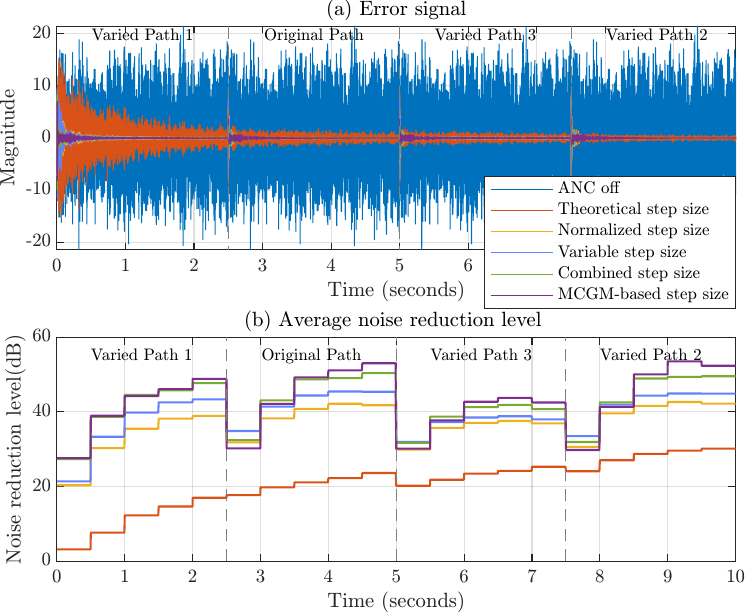}
    \caption{Noise reduction performance of FxLMS with different step-size strategies under the varying secondary path: (a) the error signal and (b) the average noise reduction level.}
    \label{fig:diff}
\end{figure}

Furthermore, to evaluate robustness, the secondary path was varied by $10\%$, $20\%$, and $30\%$ relative to the original. As shown in Fig.~\ref{fig:diff}, while these variations significantly degrade the FxLMS convergence, the proposed MCGM-based method consistently achieves significant noise reduction without additional computational complexity. This demonstrates the strategy's adaptability to dynamic acoustic environments.

%---------------------------------------------------------------------------------------
%---------------------------------------------------------------------------------------

\section{Conclusion}\label{sec:conlusion}

% To address the slow convergence problem of the FxLMS algorithm in ANC systems, numerous variable step-size strategies have been proposed, but they typically demand higher computational resources and exhibit limited adaptability across different noise conditions. In this paper, a novel MCGM-based approach is proposed to determine an appropriate step size from the noise dataset, thereby accelerating FxLMS convergence without increasing computational burden. Moreover, a forgetting factor is incorporated to mitigate the initialization effects. Simulation results confirm its robustness and superior noise reduction performance across broadband noise and multiple real-world noise types, demonstrating its potential for practical ANC applications.

To address the slow convergence issue of the filtered-reference least mean square (FxLMS) algorithm in active noise control (ANC) systems, numerous variable step-size methods have been proposed; however, these typically require higher computational resources and exhibit limited adaptability under varying noise conditions. In this paper, a novel Monte Carlo gradient meta-learning (MCGM)-based approach is proposed to determine an appropriate step size using the noise dataset—thereby accelerating FxLMS convergence without imposing additional computational burden. Furthermore, a forgetting factor is integrated to mitigate initialization-related effects. Simulation results validate the approach’s robustness and superior performance for broadband noise, multiple real-world noise types, and varying secondary path, highlighting its potential for practical ANC implementations.

\vfill\pagebreak

\bibliographystyle{IEEEbib}
\bibliography{refer}

\end{document}